
\input harvmac.tex
\hfuzz 15pt
\input amssym.def
\input amssym.tex
\input epsf\def\tfig#1{{
\xdef#1{Fig.\thinspace\the\figno}}Fig.\thinspace\the\figno
\global\advance\figno by1}


\input epsf

%


\def\({\big(}
\def\){\big)}

\def\inv{^{-1}}

 \def\frac#1#2{ {{\textstyle{#1\over#2}}}}
\def\inv{^{\raise.15ex\hbox{${\scriptscriptstyle -}$}\kern-.05em 1}}

\def\a{\alpha}
\def\b{\beta}
\def\g{\gamma}
\def\d{\delta}

\def\l{\lambda}

\def\s{\sigma}

\def\G{\Gamma}
\def\D{\Delta}

\def\O{\Omega}

\def\la{\langle} \def\ra{\rangle}

\def\IR{{ \Bbb R} }
\def\IZ{{ \Bbb Z} }

\def\dC{C\kern-6.5pt I}

\def\CG{{\cal G}}              \def\CI{{\cal I}}
              
       \def\CN{{\cal N}}       
              \def\CR{{\cal R}}
              
\def\CV{{\cal V}}

\def\CV{{\cal V}}

\def\bz{\bar z}

\input amssym.def
\input amssym.tex
\def\IZ{\Bbb Z}\def\IR{\Bbb R}


 \def\za{\alpha} \def\zb{\beta}

  \def\zs{\sigma}

  \def\zT{\Theta} 
 \def\H_{H_{1,2}}  \def\zT_{\Theta_{1,2}}
 \def\O_{O_{1,2}} \def\bH_{{\bar H}_{1,2}} 
  
 \def\V_{V_{1,2}} 
 \def\D_{D_{1,2}} \def\bD_{{\bar D}_{1,2}}


       \def\cD{{\cal D}}

     \def\cD_{{\cal D}_{1,2}} 
 \def\bcD_{{\bar {\cal D}}_{1,2}}


   \def\dC{I\!\!\!\!C}

\def\N{\CN}




\def\dal{
\vbox{
\halign to5pt{\strut##&
\hfil ## \hfil \cr
&$\kern -0.5pt
\sqcap$ \cr
\noalign{\kern -5pt
\hrule}
}}\ }



\font\maju=cmtcsc10

\def\nimreps{{{\maju nim}}-reps}

\def\zg{\gamma}
\def\bi{\bar{i}} \def\bj{\bar{j}}
\def\fqd{d_{-\frac{b}{2}}}

\def\un{{\bf 1}}
\def\ket{\rangle}
\def\bi{\bar i}
\def\bj{\bar j}

\def\tn{\tilde n}

\def\tn{\tilde n}

\def\tN{{\widetilde N}}
\def\tV{{\widetilde V}}


\def\cmp#1#2#3{{\it Comm. Math. Phys.} {\bf  #1} (#2) #3}

\def\jpa#1#2#3{{\it J. Phys.} {\bf A #1} (#2) #3}

\def\npb#1#2#3{{\it Nucl. Phys.} {\bf B #1} (#2) #3}
\def\npbfs#1#2#3#4{{\it Nucl. Phys.} {\bf B #1} [FS#2] (#3) #4}
\def\plb#1#2#3{{\it Phys. Lett.} {\bf B #1} (#2) #3}

\def\hepth#1{{\tt hep-th/#1}}


%
\lref\FF{B.L. Feigin and D.B. Fuks,  {\it Casimir operators in modules over Virasoro  algebra}, {\it Sov. Math. Dokl. } {\bf 27} (1983) 465.}
\lref\MS{G. Moore and N. Seiberg, 
{\it Classical and quantum conformal field theory}, 
\cmp{123}{1989}{177}.}

\lref\MSL{G. Moore and N. Seiberg, 
  {\it Lectures on RCFT,
 physics, geometry and topology}, Plenum Press, New York, USA (1990).}
\lref\Verlinde{E. Verlinde, {\it Fusion rules and modular transformations in conformal field theory},
 \npbfs{300}{22}{1988}{360}.
 }
\lref\PasM{V. Pasquier, 
{\it Operator content of the ADE lattice models}, \jpa{20}{1987}{5707}.
}
\lref\PTN{
B. Ponsot and  J. Teschner,  {\it Liouville bootstrap via harmonic analysis on a noncompact quantum group}, 
\hepth{9911110}.}
\lref\PTNa{B. Ponsot and  J. Teschner,  {\it Clebsch-Gordan and Racah-Wigner coefficients for a continuous series of representations of $U_q(sl(2,\CR))$}, \cmp{224}{2001}{613},
arXiv:math/0007097.}
\lref\PTbd{B. Ponsot and  J.Teschner, 
{\it Boundary Liouville Field Theory: Boundary three point function}, 
 \npb{622}{2002}{309},
 \hepth{0110244}.}
 \lref\BPb{B. Ponsot, {\it Remarks on the bulk-boundary structure constant in Liouville field theory}  (2005), unpublished preliminary manuscript (private communication). }
\lref\BPa{B. Ponsot, {\it Liouville theory on the pseudosphere: Bulk-boundary structure constant}, 
\plb{588}{2004}{105-110},  \hepth{0309211}.}
\lref\Uen{M. Nishizawa and K. Ueno, {\it Integral solutions of q-difference equations of the hypergeometric type with $|q|=1$}, q-alg/9612014.}
\lref\PZt{V.B. Petkova and J.-B. Zuber,
 {\it Generalised twisted partition functions},
\plb{504}{2001}{157}, 
\hepth{0011021}.}
\lref\BPPZ{R.E. Behrend, P.A. Pearce, V.B. Petkova and J.-B.  Zuber, 
 {\it Boundary conditions in rational conformal theories},
 \npb{579}{2000}{707},
 \hepth{9908036}. }
\lref\PZ{V.B. Petkova and J.-B. Zuber, 
{\it The many faces of Ocneanu cells},
\npb{603}{2001}{449},
 \hepth{0101151}v3.}
\lref\PZa{V.B. Petkova and J.-B. Zuber, 
 {\it On structure constants of $sl(2)$ theories},
\npb{438}{1995}{347},
\hepth{9410209}.}
\lref\Cardy{J.L. Cardy, {\it Boundary conditions, fusion rules and the Verlinde formula},
\npb{324}{1989}{581}.}
%
\lref\Lew{D.C. Lewellen, 
{\it Sewing constraints for conformal field theories on surfaces with boundaries}, 
\npb{372}{1992}{654}.
}
\lref\CarL{J.L. Cardy and D.C. Lewellen, 
{\it Bulk and boundary operators in conformal field theory}, 
\plb{259}{1991}{274}.
}
\lref\Run{I. Runkel, 
{\it Boundary structure constants for the A-series Virasoro minimal models}, 
\npb{549}{1999}{563},
 \hepth{9811178}.}
 \lref\RunP{I. Runkel,
{\it Perturbed defects and T-systems in conformal field theory},
{\it  J. Phys.} {\bf A 41} (2008) 105401, arXiv:0711.0102.}
\lref\AO{A. Ocneanu,  
{\it Paths on Coxeter diagrams: From Platonic solids and singularities to minimal models 
and subfactors}  (Notes recorded  by  S. Goto),  in {\it Lectures on operator theory}, 
ed.  B. V. Rajarama Bhat 
et al.,  Fields Institute Monographs, 
AMS Publications  (2000) 243.}
%
 \lref\ZZ{A. Zamolodchikov  and Al.  Zamolodchikov,  
 {\it Liouville field theory on a pseudosphere},
 \hepth{0101152}. }
 \lref\FZZ{ V.  Fateev, A.  Zamolodchikov  and Al.  Zamolodchikov,
 {\it Boundary Liouville field theory I. Boundary state and boundary two-point function}, 
 hep-th/0001012. }
   \lref\KacC{V.G.  Kac, {\it Laplace operators in modules of infinite-dimensional  Lie algebras and theta functions},  {\it Proc. Natl. Acad. Sci.  USA}  {\bf 81} (1984) 645.}
\lref\FFRSd{
 J. Fr\"ohlich, J. Fuchs, I. Runkel and  C. Schweigert,  {\it Kramers-Wannier duality from conformal defects}, {\it Phys. Rev. Lett.}  {\bf 93} (2004) 07061, cond-mat/0404051.}
 \lref\FFRSt{J. Fr\"ohlich, J. Fuchs, I. Runkel and  Ch. Schweigert,
 {\it Duality and defects in rational conformal field theory},  \npb{763}{2007}{354},
           \hepth{0607247}.}
\lref\BaGa{ C.  Bachas and M.  Gaberdiel,
{\it Loop operators and the Kondo problem}, 
 {\it JHEP} {\bf 0411} (2004) 065, \hepth{0411067}.}              
\lref\Alekseev{A.  Alekseev and S. Monnier, 
{\it Quantization of Wilson loops in Wess-Zumino-Witten models}, 
 {\it JHEP} {\bf 0708} (2007) 039, \hepth{0702174}.}        
\lref\hosomichi{K.~Hosomichi, {\it Bulk-boundary propagator in Liouville
theory on a disc}, {\it JHEP}  {\bf 0111}  (2001) 044, \hepth{0108093}.  }
\lref\BPa{B. Ponsot, {\it Liouville theory on the pseudosphere: Bulk-boundary structure constant}, 
\plb{588}{2004}{105},
 \hepth{0309211}.}
\lref\Sar{G. Sarkissian,  {\it Defect and permutation branes in the Liouville field theory}, 
 \npb{821}{2009}{607}, 
 arXiv: 0903.4422. }
\lref\DGOT{N. Drukker, J. Gomis, T. Okuda and  J.  Teschner, 
{\it Gauge theory loop operators and Liouville theory}, 
 arXiv:0909.1105v3.}
\lref\AGT{ L.F. Alday, D. Gaiotto and  Yu. Tachikawa, {\it Liouville correlation functions from four-dimensional gauge theories},  
{\it Lett. Math. Phys.} {\bf 91} (2010) 167, 
arXiv:0906.3219.}
\lref\AGGTV{L.F. Alday, D. Gaiotto, S. Gukov, Yu. Tachikawa  and H. Verlinde, 
{\it Loop and surface operators in ${\cal N}=2$ gauge theory and Liouville modular geometry}, 
arXiv:0909.0945v2.}
\lref\Gomis{J. Gomis, {\it Loops and defects in 4D gauge theories and 2D CFT's}, 
talk given at the ASC  {\it Workshop  on Interfaces and Wall-crossings}, Nov 30-Dec 4 (2009),  Munich, Germany.}
\lref\FPS{ P. Furlan, V.B. Petkova and M. Stanishkov,  
{\it Non-critical string  pentagon equations and their solutions}, 
\jpa{42}{2009}{304016}, arXiv:0805.0134.}


\overfullrule=0pt
\Title{\vbox{\baselineskip12pt\hbox
{}\hbox{}}}
{\vbox{\centerline
 {On the crossing relation in the presence}
 \medskip
 \centerline{ 
  of defects}
 \vskip 0.8cm
 \bigskip
  \vskip2pt
}}


 \centerline{ V.B. Petkova}
\vskip 5pt
\medskip
\bigskip
   \centerline{ \vbox{\baselineskip12pt\hbox
{\it  
Institute for Nuclear Research and Nuclear Energy, }
}}
 \centerline{ \vbox{\baselineskip12pt\hbox
 {\it Bulgarian Academy of Sciences, 
 }
 }}
 \centerline{ \vbox{\baselineskip12pt\hbox
 {\it 72 Tzarigradsko chaussee, 
 1784 Sofia, Bulgaria
 }
 }}

\vskip 2.5cm
 

\noindent
The OPE of local operators in the presence of defect lines is considered
both in the rational CFT and the $c>25$ Virasoro  (Liouville) theory.
The duality transformation  of  the 4-point  function   with inserted defect  
operators  is explicitly computed. The two channels of the correlator  
reproduce  the  expectation values of the  Wilson and 't Hooft operators, 
recently discussed in  Liouville theory  in relation to the  AGT conjecture. 

\bigskip
\bigskip\bigskip

\noindent

 \vskip 1.5cm
 
 \noindent
e-mail: petkova@inrne.bas.bg 
\Date{}


\newsec{Introduction}


\subsec{\bf The problem}

Topological defects are defined algebraically as operators commuting with
the left and right  copies of the chiral algebra \PZt, and in particular are invariant under diffeomorphisms,
\eqn\commut
{ [L_n,X]=[\bar{L}_n,X]=0 \,.  
 }
 We are concerned in this note  with the 4-point crossing relation in the presence of defect 
 operators:
 for trivial defects  it reduces to the standard Belavin-Polyakov-Zamolodchikov
  duality relation for the correlators of local 2d fields.
The presence of  a defect line  inserted between two local  operators
   modifies their  operator product expansion (OPE),  resulting in creation of defect fields,
and one is  interested in the computation of the corresponding OPE coefficients. 
A special case of this extended BPZ relation has been exploited  in \PZ\    to derive
  a general  formula for the relative  coefficients of  the OPE of local fields of integer spin in the rational
  non-diagonal theories.  Here we restrict to the diagonal theories where
 the computation of the duality transformation in the presence of defects
  is a straightforward consequence of two basic identities in CFT:  the pentagon identity for
   the quantum 6j symbols (the fusing matrices)
and the Moore-Seiberg \MS, \MSL\ torus identity, an equation for the 1-point modular matrix.

 As a  side result of this computation,  extended to the non-rational $c>25$
 Virasoro theory (Liouville CFT),  one
 obtains an explicit general expression for the expectation value of
 the 't Hooft loop operator in  Liouville theory,
 defined as the dual to that of the Wilson loop operator.
 The formula essentially reproduces
  the recently proposed ad hoc expression  
  \AGGTV, \DGOT\  and thus confirms the assumed duality of the two operators. 

\medskip
The effect of some  defect lines as creating "disorder" fields when attached to the local fields was pointed
out  already  in \PZ. It was later thoroughly  analysed in \FFRSd\ and in particular the precise conditions on the type of the defect fusion algebra leading to Kramers-Wanniers  type  duality in the rational case were described; for more on the defect fields from a TFT point of view see   \FFRSt, \RunP.  A topological defect interpretation of the Wilson loop operator  in the 2d rational theories has been discussed e.g. in  \BaGa, \Alekseev.  The
construction of the  loop operators 
 in \AGGTV,  \DGOT\ 
 does not refer to defects.\foot{The present  work was in an advanced stage
 when
 it was announced
 \Gomis\ that  a parallel work on the  defect  interpretation of the construction in  \AGGTV,  \DGOT\ is under way.}


 \subsec{\bf Preliminaries on  the topological defects}
 

Let us summarise some of the consequences of the  definition \commut\ studied in \PZt, \PZ.
In a  rational CFT with a set $\{\CI\ni j\}$ of representations the solutions of \commut\ read
\eqn\lincom
{X_x= \sum_{j,\bj}\sum_{\za,\za'=1, \dots Z_{j,\bar{j}}} {\Psi_x^{(j,\bj;\za,\za')}\over\sqrt{S_{1j}
S_{1\bj}}}\,  P^{(j,\bj;\za,\za')}\ ,}
where $P^{(j,\bj;\za,\za')}$ are projectors in the  representation spaces  $ (\CV_j \otimes  \overline{\CV}_{\bj })^{\b}\,,
$ and the sum is restricted  to pairs $(j,\bj)$, allowed by the nonzero values of the given modular invariant  matrix  $Z_{j,\bar{j}}$, taken with their multiplicity $ \b=1,\dots Z_{j,\bar{j}} $;
$\Psi$ is a unitary matrix of size $\sum_{j,\bar{j}} Z^2_{j\bar{j}}$.
As in the  computation of the cylinder partition function, which
 leads to non-negative integer matrix representations
(\nimreps)  $n_{ja}{}^b  
$ of the Verlinde algebra \Cardy, \BPPZ\
\eqn\nmrep{\eqalign{
n_i n_j& =  
\sum_{s}\N_{ij}{}^s n_s\,,   \qquad {}^Tn_j= n_{j^*}\,, \ \  \,  j\in \CI \cr
\Leftrightarrow & \ n_{ja}{}^b =\sum_{l\in \CI,\a=1,\dots Z_{ll}} {S_{jl}\over S_{1l} }
\psi_a^{(l,\a)}\, \psi_b^{(l,\a)\, *} \,, 
\psi \psi^\dagger =Id= \psi^\dagger \psi\,,
}}
one constructs partition functions on the torus
$Z_{x_n| x_1 x_2 \dots x_{n-1}}$ (or,   on the cylinder $Z_{b| x_1 x_2\dots x_n\, a}$),  inserting arbitrary number of defect operators.
  They are sesquilinear (respectively linear) combinations of the characters of the chiral algebra representations with non-negative integer coefficients  $\tV_{i \bar{j}; x_1 \dots x_{n-1}}{}^{x_n}\,, i,\bar{j}\in \CI$
(resp $n_{j; x_1 \dots x_n; a}{}^b$). 
 The case of two defects on the torus
 leads
  to an equation for the multiplicities  analogous  to \nmrep\
\eqn\doublefus{\eqalign{
 \tilde{V}_{i_1j_1} \tilde{V}_{i_2j_2}&=
\sum_{i_3,j_3} \N_{i_1i_2}{}^{i_3} \N_{j_1j_2}{}^{j_3}\
\tilde{V}_{i_3j_3}\,, \quad 
\tV_{ij^*; 1}{}^1=Z_{ij}\,, \quad {}^T\tV_{ij}= \tV_{i^*j^*}\cr
\Leftrightarrow  \ \tV_{ij; x}{}^y&=
\sum_{l,\bar{l},\za,\za'}\,
{S_{il}S_{j\bar{l}} \over S_{1l}S_{1\bar{l}}}\
\Psi_x^{(l,\bar{l}; \za,\za')}\ \Psi_y^{(l,\bar{l};\za,\za')\, *} \,.
}}
The classification  of the topological defects amounts in the classification of the
NIM-reps  \doublefus. In
  the sl(2)  related  cases  it confirms the results of Ocneanu, visualised  by his generalised ADE diagrams, with vertices  associated  with the set of defects \AO.
Another distinguished set of non-negative integers $\tV_{1 1; y x }{}^{z}=\tN_{yx}{}^z$  is provided by the identity contribution  of  left and right characters in the torus partition function with three inserted defects, $Z_{z| y\, x}$. This set serves as structure constants of an associative, in general non-commutative,  algebra,    the fusion algebra of defects (=Ocneanu graph algebra), as it allows to compute the fusion
of two defects
\eqn\tildN{\eqalign{
X_y X_x & = \sum_z \tN_{yx}{}^z\, X_z\,.
\cr
}}
The action of the defects on the boundary states
 \eqn\bulkact{\eqalign{
 X_x | a \ra &
 =\sum_c \tilde{n}_{ a x}{}^c | c \ra\,, \cr
\tn_{ax}{}^c&=\sum_{j,\a,\b}\psi_a^{(j,\a)}{\Psi_x^{(j,j;\a,\b)}
\over\sqrt{S_{1j} S_{1\bj}}
}\psi_c^{(j,\b)*} \,, \ \a,\b=1,\dots Z_{j,j}
}}
 introduces another set  of non-negative integers $\{\tilde{n}_{ a x}{}^c\}$,   interpreted
 as the multiplicities $n_{1; x; a}{}^c=\tilde{n}_{ a x}{}^c$ of the identity character
 in the cylinder partition functions in the presence of one  defect, $Z_{b;x, a}=\sum_j n_{j; x; a}{}^b\chi_j(\tau)= \sum_j (\tn_x n_j)_a{}^b \chi_j(\tau)$.
Combining 
 relations \tildN\ and \bulkact\  implies that
this set of matrices provides
NIM-reps of the Ocneanu algebra
\eqn\tildn{
\tn_x \tn_y = \sum_z \tN_{xy}{}^z \tn_z  \,.
}
Thus the computation of  the partition functions on the torus and the cylinder with an arbitrary
number of defects inserted is  reduced to the knowledge of several basic structure constants,
or, equivalently, the knowledge of the sets of unitary matrices $\{\psi, \Psi\}$ in \nmrep, \doublefus, e.g.,
\eqn\simpld{
\tV_{ij^*; x}{}^z= \sum_y \tN_{x y}{}^z\, \tV_{ij^*;1}{}^y\,,\ \   \tV_{ij^*; x_1 x_2}{}^z = \sum_u \tN_{x_1 x_2}{}^u\, \tV_{ij^*;u}{}^z\,.
}
 In the field  interpretation, these 
multiplicities encode the possible  holomorphic - antiholomorphic
content $(i,j)$  of defect
fields; in general  they correspond to non-local
2d fields.

In the case of a "diagonal" theory, i.e., described by a modular matrix with $Z_{j\bj}=\delta_{j\bj}$,
the sets of defects and boundaries can be identified with the set $\CI$  of representations of the chiral algebra
 and
 $n,\tn, \tN$ coincide with the Verlinde
 multiplicities $\N$, while
  \eqn\dig{
\tV_{ij}=\N_i\N_j\,, \ \  \tV_{ij;1}{}^y =\N_{i j}{}^y\,,
 }
 etc.
Accordingly
the matrices diagonalising these multiplicities  reduce to the  modular matrix,
 $\Psi_x^{(j,\bar{j});\a,\b} = \delta_{j \bar{j}} \delta_{\a 1} \delta_{\b 1}S_{xj}$,
and   the eigenvalue in the r.h.s. of \lincom\  is expressed by the ratio of modular matrix  elements  ${S_{xj}/ S_{1j}}$.    In the diagonal case the action of the defect operator $X_x$ (on $(\CV_j \otimes  \overline{\CV}_{j}$) coincides with the action of its  chiral analog acting on %
  $\CV_j$ (one to one with the Ishibashi states)
    \eqn\chdef{X_x^I
    =\sum_{j} {S_{xj}\over S_{1j}} \sum_k |j,k\ra \la j,k|
    }
In the WZW case this operator
  can be identified \Alekseev\  with  a
   generalised Casimir operator  \KacC (see also \FF\ for the Virasoro minimal models)
   giving precise meaning of the
 Wilson loop operator \BaGa.
  The twisted partition function in the diagonal case can be interpreted alternatively in terms
  of related  operators, associated with the two cycles of the torus
  \eqn\chact{\eqalign{
  \hat{X}_x
  (a) \chi_p(\tilde{\tau})&= {\rm Tr}_p(e^{2 i\pi  \tilde{\tau}(L_0- c/24)}X_x^{I}) = {S_{xp}\over S_{1p}}  \chi_p(\tilde{\tau})\cr
\hat{X}_x
(b)
\chi_p(\tau)&= \sum_i S_{pi }  \hat{X}_x^{I}(a)  \chi_i(\tilde{\tau})=\sum_i S_{pi }  {S_{xi}\over S_{1i}} \chi_i(\tilde{\tau})= \sum_s \N_{xp}{}^s \chi_s(\tau)
}}
This reproduces the monodromy operators in the derivation of the Verlinde formula \Verlinde.

   In the Liouville case the sum in \lincom\ is replaced by an integral
  over a continuous  series of representations. 
   The
  modular matrices were computed
  in \ZZ\ in the two basic cases - with the second representation $x$  also belonging  to the continuous principal series,   
  or,  with $x$ belonging to the $c>25$ infinite discrete degenerate series. The latter is  parametrised by a pair $(m,n)$ of positive integers, i.e., $x$ labels  a factor representation of a degenerate Verma module of  scaling dimension $\triangle(x)= 
  x(Q-x)$,  with  $x= x_{m n}=-\frac{m-1}{2} b -\frac{n-1}{2 b}\, , $ 
   $Q= b+\frac{1}{b}$. Thus,  there are two types of defects in the Liouville theory, with two different  ratios of modular matrix elements in \lincom,
   corresponding  to the
  two types  FZZ \FZZ\ or ZZ \ZZ, respectively, of boundary states;   this has been discussed in detail recently in \Sar.

For the purpose of the comparison with  recent work  \AGGTV, \DGOT\  on the relation of  Liouville theory to  the 4d supersymmetric gauge theories \AGT,
we perform in more detail also the explicit computation in the Liouville 
case. We will restrict mostly to the ZZ case, a quasi-rational theory,  
similar in many respects to the rational $c<1$  theory.
 The main section  2 deals with the diagonal  rational case.  The technical difference with the Liouville theory  is in the use of different normalisation of the chiral vertex operators  - traditionally in the minimal models the $\IZ_2$ symmetry is effectively fixed, the reflected operators are identified, and in particular one can use bases of conformal blocks for which the  fusing matrices are unitary.   The three parts of  section 3   deal with the Liouville case and contain:  a collection of  basic   formulae 
   (sect. 3.1);  more details on  the Liouville    torus identity and the crossing relation in the Liouville case (sect. 3.2);   an explicit example compared   with the proposed expression for the expectation value of the 't Hooft operator in \AGGTV, \DGOT\ (sect. 3.3). The Appendix recalls the OPE formula of \PZ.


 \newsec{The duality relation in the rational case}
 
 
For simplicity of notation 
 we shall restrict to the $sl(2)$ case, but will keep the  conjugation so that
 the higher rank generalisation is straightforward.  In the rational case  one can choose a unitary fusing matrix $F$.
 
  As in \PZ\ we consider 
 a $4$-point function of four local fields with insertion of two  defects; for the purpose here we shall
 restrict to the diagonal case with scalar fields; the second label will be suppressed, 
\def\tz{\tilde{z}}
\def\ket{\rangle}
\def\bk{\bar{k}}
\eqn\vone {\eqalign{
G&=\la 0|\Phi_{a_4}(\tilde{z}_4, \bar{\tilde{z}}_4)\,
 \Phi_{(a_3}(\tz_3, \bar{\tz}_3)\, 
 X_x\,
\Phi_{a_2}(\tz_2, \bar{\tz}_2)\,
\Phi_{a_1}(\tz_1, \bar{\tz}_1)\, X_x
 |0\ket \cr 
&= \sum_j {S_{xj}\over S_{1j}}  {S_{x 1}\over S_{11}}|\CG_{j}(a_4, a_3, a_2, a_1;\tz)|^2\cr
&=d_x \sum_{\g,\d} \sum_j {S_{x j}\over S_{1j}} 
 F_{j  \g}\left[ {a_4\atop a_3^*}{a_1\atop a_2}\right]  F_{j  \d}^*\left[ {a_4\atop a_3^*}{a_1\atop a_2}\right]\,
 \CG_\g(a_3, a_2, a_1, a_4;z)\CG_\d(a_3, a_2, a_1,a_4;\bz)\,.
}}
In the second line we have used that in this channel the defects are  diagonalised,  applying the OPE expansion of the local fields
and  inserting the diagonal version of the general formula for the 2-point function
\eqn\deftw{
\la 0|\Phi_{(J^*,\a)} X_x \Phi_{(J',\b)}|0\ra = \delta_{j,j'}\delta_{\bar{j}, \bar{j'}}{\Psi_{x}^{(J;\a,\b)}\over \Psi_1^{J}}\la 0|\Phi_{(J^*,\a)}  \Phi_{(J,\b)}|0\ra \,.
}
In the third line of \vone\  we have performed the braiding of the chiral blocks. In the new channel 
 the defect  modifies the OPE expansions of two  local fields
\eqn\voneb{\eqalign{
&G=\la 0|
 \Phi_{(a_3}(z_3, \bz_3)\, 
 X_x\,
\Phi_{a_2}(z_2, \bz_2)\,
\Phi_{a_1}(z_1, \bz_1)\, X_x
\Phi_{a_4}(z_4, \bz_4)|0\ket \,.
}}
It remains  to perform the initial summation over $j$ in  \vone, i.e., 
\eqn\derrat{\eqalign{
&A_{\g, \d}^{(x)}=\sum_j {S_{x j}\over S_{1j}} 
 F_{j  \g}\left[ {a_4\atop a_3^*}{a_1\atop a_2}\right]  F_{j  \d}^*\left[ {a_4\atop a_3^*}{a_1\atop a_2}\right]\,, \ \ {\rm with} \  A_{\g, \d}^{(1)}=\delta_{\g\d} \,.
 }}
The first  step is, using a standard relation derived from the pentagon identity, to  rewrite  the product of F matrices \derrat\ as 
 \eqn\derrata{
 F_{j  \g}\left[ {a_4\atop a_3^*}{a_1\atop a_2}\right]  F_{j  \d}^*\left[ {a_4\atop a_3^*}{a_1\atop a_2}\right]= 
  d_{j}\sqrt{d_\g d_\d \over d_{a_1}d_{a_2} d_{a_3} d_{a_4}}\, F_{a_1a_3^*}\left[ {a_2\atop j}{\g \atop a_4^*}  \right]  F_{a_4^* a_2}\left[ {a_3^*\atop j }{\d^* \atop a_1} \right]\,.
 }
 We have also used  the relation derived from the pentagon identity,  taking  into account 
 the symmetries of the 6j symbols  and using that $F_{1 1 }\left[ {a\atop a}{a^* \atop a}\right]=1/d_a$:
 \eqn\qdimg{
 F_{1 m}\left[ {j\atop j}{i \atop i^*}\right] F_{m 1}\left[ {i^*\atop j}{i \atop j}\right] ={d_m\over d_i d_j}\, \N_{ji}{}^m\,.
 }
It implies that for a unitary $F$ the values of $F_{1 c}$ and $F_{y 1} $ are determined up to a sign
 in terms of a square root of a ratio of q-dimensions;
 the positive sign is chosen. 
 We next   apply the pentagon identity itself,  representing the product in \derrata\ as   a sum 
  \eqn\derratb{\eqalign{
 &  F_{a_4^* a_2}\left[ {a_3^*\atop j}{\d^* \atop a_1} \right] F_{a_1a_3^*}\left[ {a_2\atop j}{\g \atop a_4^*}
  \right]  = \sum_y
   F_{y a_3^*}\left[ {a_2\atop a_2}{\g \atop \d^*}\right]   F_{a_4^*  y}\left[ {\g\atop a_1}{\d^* \atop a_1}\right] 
     F_{a_1 a_2}\left[ {a_2\atop j}{y \atop a_1}\right] \cr
   &=  \sum_y F_{ a_3^* y^*}\left[ {\g^*\atop \a_2}{\d \atop a_2}\right]   F_{a_4^*  y}\left[ {\g\atop a_1}{\d^* \atop a_1}\right] \,   F_{y^* j}\left[ {a_2\atop a_2}{a_1 \atop a_1^*}\right] {F_{a_1^* 1}\left[ {j \atop a_2}{j ^* \atop a_2}\right] \over   F_{y^* 1}\left[ {a_2\atop a_2}{a_2^* \atop a_2}\right] }\,.
}}
 Inserting this into \derrat\ we can now perform the  summation over $j$,   using 
 the  MS  torus identity \MS, \MSL. It can be written e.g. as
 \eqn\tidar{\eqalign{
&\sum_s \, S_{r i}(s)\, 
F_{qs}\left[ {j_1\atop r}{j_2\atop r }\right] \sum_m \, e^{2  \pi i (\triangle_{i}\!-\!\triangle_p\!+\!\triangle_{j_1}\!+\!\triangle_{j_2}\!-\!\triangle_m)} F_{s m}\left[{i \atop i}  {j_2\atop j_1} \right] F_{m p}\left[ {j_1\atop i}{j_2 \atop i}\right] \cr
&= S_{q i}(p)
F_{r p}\left[ {j_2\atop q}{j_1 \atop q}\right] \,.
}}
Taking $r=1$, hence $s=1$, gives an expression for the 
 modular matrix $S(p)$ of one point functions on the torus 
   \eqn\toraar{\eqalign{
S_{ji}(p)&=
{S_{1i}\over F_{1p}\left[ {j\atop j}{j^* \atop j}\right] } 
\sum_m\, e^{i \pi (2(\triangle_i\!+\!\triangle_j\!-\!\triangle_m)\!-\!\triangle_p)}F_{1 m}\left[ {j\atop j}{i \atop i^*}\right] F_{m p^*}\left[ {i^*\atop j}{i \atop j}\right]\,, \cr
}}
and $
S_{ji}^*(p)= e^{i \pi \triangle_p} S_{j^*i}(p^*)$.
 Inserting the transposed version of \toraar\  in \tidar\  -   taken for      $p=1$,  
and inverting 
 with $F_{qs}^{-1}$ gives a  "fusion" like representation for the product of two S-matrices, 
   \eqn\derratc{\eqalign{
 & 
{d_x\over  F_{y^*1}\left[ {a_2\atop a_2}{a_2^* \atop a_2}\right] } 
\sum_j 
 {S_{x j}\over S_{11}} F_{y^* j}\left[ {a_2\atop a_2}{a_1 \atop a_1^*}\right]    F_{a_1^* 1}\left[ {j \atop a_2}{j^* \atop a_2}\right] 
   = 
    {S_{a_2 x}(y^*)\over S_{11}} e^{i\pi \triangle_y}
 {S_{a_1 x}(y)\over S_{11}}\,.
 }}
 In particular,  for $y=1$  \derratc\  reproduces, using \qdimg,  the Verlinde formula for the fusion multiplicity 
 $\N$.
Combining \derrata, \derratb, \derratc\ we get finally for \derrat, 
\eqn\derratd{\eqalign{ 
d_x\, A_{\g, \d}^{(x)}
&=\sqrt{d_\g d_\d \over d_{a_1}d_{a_2} d_{a_3} d_{a_4}}\
 \sum_y   
F_{ a_3 y}^*\left[ {\g\atop a_2^*}{\d^* \atop a_2^*}\right] {S_{a_2^* x}^*(y)\over S_{11}}{S_{a_1 x}(y)\over S_{11}}F_{a_4^*  y}\left[ {\g\atop a_1}{\d^* \atop a_1}\right] \,.
}}
The range of $y$ in \derratd\ is determined by the    multiplicities 
$\N_{xy}{}^x\,, \N_{a y}{}^{a}$  of the 1-point blocks
on the torus, transformed by the 1-point modular matrix $S_{a x}(y)$, $a=a_1,a_2$. We see that we can interpret $y$ as a 
defect label, appearing in the defect fusion multiplicity $\N_{xy}{}^x$. 
In turn the values of $y$ restrict  the possible 
pairs  $(\g, \d)$ of  representations labelling   the left and right conformal blocks in \vone\ as dictated by the  
multiplicity $\N_{\g \d^*}{}^{y}$, implicit in the $F$  matrices.
Altogether this corresponds to the first of the relations in \simpld\ for $z=x$,
\eqn\simlda{
\tV_{\g \d^*; x}{}^x = \sum_y \tN_{xy}{}^x \tV_{\g,\d^*;1}{}^y= \sum_y  \N_{xy}{}^x \N_{\g \d^*}{}^y\,.
}
In other words,   $\tV_{\g, \d^*;1}{}^y$ is the  multiplicity  of a defect field ${}^y\Phi_{(\g,\d)}(z,\bar{z})$ that  is created in the 
OPE of  two local operators,  modified by the inserted
defect line operator $X_x$ as in \voneb, 
\eqn\feop{\eqalign{
& \Phi_{a_1}(z_1,\bar z_1) X_x \Phi_{a_4}(z_4,\bar z_4)=z_{14}^{-\triangle_{ij}^\g} \bar{z}_{14}^{-\triangle_{ij}^\d} 
 \sum_{\g, \delta, y} d_{a_1,a_4; x}{}^{(\g, \delta); y}\
{}^y\Phi^{}_{(\g, \d)}(z_2,\bz_2) +... }}
 and the OPE coefficients in \feop\ are determined from  \derratd\ up to a normalisation of the defect field 2-point function. 
 Alternatively, inserting
$$
\d_{y,y'}= \sum_q F_{y q}\left[ {x\atop x}{\d^* \atop \g}\right] 
F_{y' q}^*\left[ {x\atop x}{\d^* \atop \g}\right] 
$$
we can rewrite \derratd\ in terms of the modular matrix transforming a 2-point chiral block on the torus
\eqn\tptm{S_{i,a;x,  q}(\g_1,\g_2) = F^{-1} (S \otimes \un) F =
\sum_{y }   
 F_{y q}\left[ {x\atop x}{\g_2\atop \g_1}\right]  S_{i x}(y) F_{a y}\left[ {\g_1\atop i}{\g_2\atop i}\right] \,.
}
In the higher rank cases it depends on   four more indices, their range being determined by the  fusion multiplicities $\N_{\g_1 a}{}^i\,, \N_{\g_2 i}{}^a\, $ and $ \N_{\g_1 q}{}^x\,, \N_{ \g_2 x}{}^q$ of the chiral vertex operators involved. For \derratd\ we have
\eqn\derratt{
d_x\, A_{\g, \d}^{(x)}=\sqrt{d_\g d_\d }
\ \sum_q {S_{a_2^*, a_3; x, q}^*(\g, \d^*)\over \sqrt{d_{a_2} d_{a_3}}}
{S_{a_1, a_4^*; x, q}(\g,  \d^*)\over \sqrt{d_{a_1} d_{a_4}}} \,.
}
The summation here corresponds  to \simlda\ rewritten as
\eqn\simldb{
\tV_{\g \d^*; x}{}^x = \sum_y \tN_{xy}{}^x \tV_{\g,\d^*;1}{}^y =\sum_q \N_{\g q}{}^x \N_{\d^* x}^q\,.
}
\medskip

 Taking $\g=1= \d$ in \derratd\   represents  the leading contribution of the identity block in \vone,
 with the restrictions $a_3=a_2^*\,, a_4=a_1^*$.  
 Using \deftw\  this reproduces indeed the  r.h.s. of \derratd\ for these values. 

 In the analog of the chiral interpretation \chact\ here,  the modular transformation of the characters is replaced by braiding (fusing) of the conformal blocks on the sphere
  \eqn\chacta{\eqalign{
  \hat{X}_x(a) \CG_p(\tz)&= \la 0|\phi_{a_4, a_4^*}^1(\tz_4) \phi_{a_3 p}^{a_4^*}(\tz_3)X_x^{I} \phi_{a_2, a_1}^p(\tz_2) \phi_{a_1 0}^{a_1}(\tz_1) |0 \ra = {S_{x p} \over S_{1p}}\  \CG_{p}(\tz)\cr 
   \hat{X}_x(b)   \CG_p(z)&=  \sum_s \sum_{j} F_{ p j}^{-1}\,
 {S_{x j}\over S_{1j}}\, F_{ j s}\   \CG_s({z})= \sum_{s} A^{(x)}_{ps}  \CG_s({z})
 }}
and  formula \derratd\  gives alternative
expression for  $A^{(x)}_{ps} $.

\medskip

The  MS  torus identity  has been encountered in  the boundary CFT  in \Run, \BPPZ; it has been observed that 
 in the diagonal case the two Cardy-Lewellen bulk-boundary eqs  \CarL, \Lew\  both originate in this identity. The first equation corresponds to \tidar\ (with boundary labels $q,r$ and $F_{r p}, F_{q s}$ substituted by the 3j symbols - the  boundary field OPE coefficients),  while the second equation 
  is identified with  the
  (transposed version of the) relation \derratc.   
  Accordingly  the  bulk-boundary  structure constant in the diagonal theory 
 is proportional to the 1-point modular matrix.  
 
 The duality transformation relating the correlators \vone\ and \voneb\
has been recently discussed in \Sar,   following   \PZ\ and comparing with the permutation brane approach;  this consideration  does not   yield,  however, 
 explicit formulae  like \derratd,  \derratt.


\newsec{The Liouville case}


\subsec{\bf Collection of Liouville formulae}


The quantum 6j symbols $F$  for the 
Liouville theory have been computed in \PTN, \PTNa. 
 It is convenient to change the normalisation 
 of the chiral blocks, ${}^{(F)}\CG_\b \to \CG_\b$
\eqn\gbl{\eqalign{
{}^{(F)}\CG_\b(\a_4,\a_3,\a_2,\a_1; \tilde{z})&=N(\a_4^*, \a_3, \b) N(\b,\a_2, \a_1) \,  \CG_\b(\a_4,\a_3,\a_2,\a_1; \tilde{z})\,, \cr
N(\b_3,\b_2,\b_1) 
&={\G_b(Q) \G_b(2\b_1)\G_b(2\b_2)\G(2Q-2\b_3)\over \G_b(2Q-\b_{123})
\G_b(\b_{12}^3)\G_b(\b_{23}^1)\G_b(\b_{13}^2)}\,, \cr
}}
($\b_{123} =\sum_i \b_i \,,\  \b_{12}^3= \b_1\!+\!\b_2\!-\!\b_3$, etc.) so that  $ \CG_\b$  transform with the matrices
\eqn\gfmat{\eqalign{
G_{\b_5
\,,\b_6}\left[\matrix{\b_3&\b_2\cr \b_4&\b_1} \right]&={N(\b_6,\b_3,\b_2) N(\b_4,\b_6, \b_1)\over N(\b_4, \b_3,\b_5)N(\s_4, \b_5,\b_1)}\,   F_{\b_5
\,,\b_6}\left[\matrix{\b_3&\b_2\cr \b_4&\b_1} \right]\,.
}}
The  Liouville bulk 3-point  constant is then given by  
\eqn\pn{\eqalign{
&
N(\b_3,\b_2,\b_1)N(Q-\b_3,Q-\b_2,Q-\b_1)
=2\pi  
\l^{-Q\over 2b} \prod_i W(Q-\b_i)\, C(\b_3,\b_2,\b_1)^{-1}\,.\cr
}}
Here $\l:=\pi \mu\,  \G(b^2)/\G(1-b^2)\, b^{2-2b^2} $ and  we have used the ZZ variable \ZZ\   
\eqn\zzw{
W(\a)=
{\G_b(2\a)\over \G_b(2\a-Q)}   
 \l^{2\a-Q\over 2b} \,
(= - 2W(iP)^{ZZ})\,.
}
Recall that 
the product of $W(\a)$ in \zzw\ and its reflected counterpart $W(Q-\a)$  is proportional to a modular matrix element,
while the ratio gives  the bulk reflection amplitude, 
\eqn\zzw{\eqalign{
W(\a) W(Q\!-\!\a)&=
{S_b(2\a)\over S_b(2\a\!-\!Q)}
 = - 4\sin \pi b(2\a\!-\!Q) \sin \frac{\pi }{b}(2\a\!-\!Q)=: S_{0\a}\, 
 \cr
{W(Q-\a) \over W(\a)}&={\Upsilon_b(2\a)\over \Upsilon_b(2\a-Q)} 
\l^{Q-2\a\over b}
=S(\a)\,.
}}
More generally, for  the case of a  degenerate representation $x$
   and generic 
   charge $\a  $ the modular matrix  reads \ZZ\ (up to an overall normalisation) 
 \eqn\modmZ{\eqalign{
  {S_{x_{m,n}\,  \alpha}}&=-4\, {\sin \pi b m (2\a\!-\!Q)} {\sin \frac{\pi n}{ b}(2\a\!-\!Q) }
   = \hat{S}_{x_{m,n}\,  \alpha}- \hat{S}_{x_{-m,n}\,  \alpha}
 }}
 where 
 \eqn\modF{
 \hat{S}_{\beta  \alpha}= 2 \cos \pi (2\a -Q)(2 \beta-Q) 
 } 
 is the  FZZ type modular matrix,
 computed   for two generic  representations.
 
The  Weyl reflected  charge in the second line of \modmZ\ corresponds to the only singular vector at  generic  $b^2$ of the
reducible Virasoro 
 module  of highest weight $\triangle(x_{m,n})$. This relation, 
coming from the character formula for the degenerate representations, 
extends to  other quantities of the theory,  e.g., 
the corresponding  fusion multiplicities,  
$\hat{\N}_{\a \b }{}^\g$ 
and $\N_{\a x_{m,n}}{}^\g$    
\eqn\fmrel{
\N_{\a\, x_{m,n} }{}^\g=\hat{\N}_{\a\, x_{m,n}}{}^\g- \hat{\N}_{\a\, x_{-m,n}}{}^\g \,.
}
Here  the l.h.s. is a  finite sum of delta functions,  while  $\hat{\N}_{\a \b }{}^\g$ is given \ZZ\ by an integral formula of  Verlinde type, i.e., it is diagonalised by $\sqrt{2} \hat{S}_{\a \d}$ in \modF\   and  its   eigenvalues (1-dimensional representations) are given by the ratios  ${\hat{S}_{\a \d}/ S_{0\d}} $.   
 
The   $F$ matrix is invariant under reflection  $\b_i\to Q-\b_i$ of any of the indices \PTN, equivalent to a complex conjugation  for pure imaginary $iP=Q-2\b$. Extended to arbitrary values of the charges, 
the gauged $G$ matrix \gfmat\ satisfies the standard
symmetry relations with the star operation understood as a reflection, $\b^*=Q-\b$
\eqn\sym{\eqalign{
G_{\b_5
\,,\b_6}\left[\matrix{\b_3&\b_2\cr \b_4&\b_1} \right] =G_{\b_5
\,,\b_6^*}\left[\matrix{\b_4^*&\b_1\cr \b_3^*&\b_2} \right] =G_{\b_5^*
\,,\b_6}\left[\matrix{\b_2&\b_3\cr \b_1^*&\b_4^*} \right] \,.
}}
The  locality of the scalar 4-point function  is rewritten  in terms of the $G$ matrices as
\eqn\loc{
\int d\g \, {S_{\b 0}\over S_{\g 0}}\,  G_{\b \g}\left[ {\a_4\atop \a_3^*}{\a_1\atop \a_2}\right]  G_{\b \g'}^*\left[ {\a_4\atop \a_3^*}{\a_1\atop \a_2}\right] = \delta(\g-\g')\,.
}
The integrals here and below  run along $ {Q\over 2} +i \IR^+$. We shall exploit the relation of 
 the fusion  matrices to the Liouville boundary field  OPE coefficients $C$ \PTbd, 
 \eqn\relbc{\eqalign{
G_{\s_2
\,,Q-\zb_3}\left[\matrix{\zb_2&\beta_1\cr \zs_3&\sigma_1} \right] ={N(Q-\b_3,\b_2,\b_1) R(\s_3,Q-\b_3, \s_1)\over R(\s_3, \b_2,\s_2)R(\s_2, \b_1,\s_1)}C_{\s_2
\,,Q-\zb_3}\left[\matrix{\zb_2&\beta_1\cr \zs_3&\sigma_1} \right] \,,
}}
where  $R$ is the ratio  of the two gauge factors; 
\eqn\rat{
R^{-1}(\s_2,\g, \s_3)={g(\s_2,\g, \s_3)\over N(\s_2,\g, \s_3)}:=
\l^{\g+\s_3-\s_2\over 2 b}{S_b(\g+\s_2-\s_3)S_b(\g+\s_3-\s_2)\over S_b(2\g)}
}
The 
OPE coefficients are related to the 
coefficients 
of the boundary field 3-point functions
 \eqn\reflC{\eqalign{
& C^{\s_3, \s_2,\s_1}_{\b_3,\b_2,\b_1}
 = C_{\s_2
\,,Q-\zb_3}\left[\matrix{\zb_2&\beta_1\cr \zs_3&\sigma_1} \right] = S(\s_3, \b_3, \s_1) C_{\s_2
\,,\zb_3}\left[\matrix{\zb_2&\beta_1\cr \zs_3&\sigma_1} \right]\,, \cr
& S(\s_3, \b, \s_1) = {g(\s_3, Q-\b, \s_1) \over g(\s_3, \b, \s_1) }\,
}} 
with the boundary reflection amplitude \FZZ\ defined in the second line.  
In the case when the three charges  $\b_i $ in \relbc\ are constrained by a charge conservation condition, the 3-point function  $C^{\s_3, \s_2,\s_1}_{\b_3,\b_2,\b_1}$
 develops poles.  The residue  corresponds to the  correlator, 
 which can be  computed in the  half-plane Coulomb gas formulation of \FZZ. We shall denote it and the 
 residues of  the corresponding   $G$ in \relbc\   by the same letters. For  $\sum_i \b_i-Q=0$ 
(absence of screening charges)  the residue is $1$,  so in these cases
$G$ reduces to the gauge factor in \relbc. 
Furthermore any  $C$ related by a reflection to a trivial one
is also simple,  being obtained by applying the boundary reflection matrix as in \reflC.
This modifies one (or two,  or three) of the  ratios \rat\ in \relbc,  replacing   $g(\s_4,\g, \s_3)$ with $g(\s_4,Q-\g, \s_3)$.  
Examples  of $G$ matrix elements  obtained this way will be  used below:  
\eqn\sgaq{\eqalign{
&G_{\b  Q}\left[ {\a^*\atop \g}{\a \atop \g}\right]  
=
{1\over d_\a}= {\sin \pi b Q \sin \frac{\pi}{b} Q \over \sin \pi b(2\a-Q) \sin \frac{\pi}{b}(2\a-Q)  }
 = 
 G_{\b  0}\left[ {\g^*\atop \a}{\g \atop \a}\right] \,,  \cr
&G_{\a \g^*}\left[ {\a^*\atop Q}{\b \atop \g}\right]=G_{\a \g}\left[ {0\atop \a}{\g \atop \b}\right]=1\,,
}}
where the quantum dimension $d_\a= {S_{0\a}\over S_{00}} $ appears; furthermore 
\eqn\ffg{\eqalign{
G_{Q+b, \a\pm  b/2}\left[ {\a\atop \a}{-\frac{b}{2} \atop Q\!+\!\frac{b}{2}}\right] & =\mp {\sin \pi b^2\over \sin \pi b(2\a-Q)}\,,\cr
G_{\a\pm b/2, Q+b}\left[\matrix{Q\!+\!\frac{b}{2}&-\frac{b}{2}\cr \a&\a} \right] &
=\pm {\sin \pi b(2\a\mp Q-Q)\over \sin \pi 2b^2}\,.
 }}
More generally, denoting 
\eqn\normsm{
G_2(\s_3, \b, \s_1) : =    {S(\s_3, \b, \s_1)\over W(Q-\b)} = S_b(2\b-Q)   {S_b(\s_{2}+\s_{1}-\b)S_b(Q-\b+\s_{2}-\s_1)\over S_b(\b+\s_{2}+\s_1-Q)S_b(\b+\s_{2}-\s_1)}
}
we can write a compact formula for the general Coulomb gas boundary coefficients $C$, obtained as a residue from the Ponsot-Teschner (PT) formula \PTbd, 
 \eqn\PTR{\eqalign{
& C_{\s_2
\,,Q-\zb_3}\left[\matrix{\zb_2&\beta_1\cr
\zs_3&\sigma_1} \right]= 2\pi \,  Res_{\b_{123}-Q+mb+n/b=0}\ C^{(PT)}_{\s_2
\,,Q-\zb_3}\left[\matrix{\zb_2&\beta_1\cr
\zs_3&\sigma_1} \right] 
 = g(Q-\b_3\,, \b_2, \b_1)^{-1} \, \times \cr
& {S_b(2\b_2+mb+\frac{n}{b})S_b(2\b_1) \over S_b(2\b_2) S_b(2\b_1+mb+\frac{n}{b})}
\sum_{k=0}^m\sum_{p=0}^n 
{G_2(\s_3\!-\!\frac{(k-m) b}{2}\!-\!\frac{p-n}{2b},Q\!-\!\b_3\!+\!\frac{(k-m) b}{2}\!+\!\frac{p-n}{2b}\,,\s_1)\over G_2(\s_3,Q-\b_3\,,\s_1)}\, \times \cr
&
{G_2(\s_3\!-\!\frac{k b}{2}\!-\!\frac{p}{2b}, Q\!-\!\b_2\!-\!\frac{k b}{2}\!-\!\frac{p}{2b}\,,\s_2)\over G_2(\s_3, Q-\b_2\,,\s_2)}{(-1)^{m(p+1)+n(k+1)+mn} \over S_b((k\!+\!1)b) S_b((m\!-\!k\!+\!1)b)S_b(\frac{p\!+\!1}{b})S_b(\frac{n\!-\!p\!+\!1}{b})}\,.
}}
Further    some of the $\b_i$  charges in \PTR\ can be set 
to degenerate values;  it is   a polynomial in the boundary 
 parameters $ 2\cos \pi b( 2\s_i-Q)\,, 2\cos \frac{\pi} {b}( 2\s_i-Q)$.  
Rewritten in terms of finite products of  sine-functions  \PTR\  admits
 analytic continuation  to the region $c<1$ \FPS\ 
 and in  this sense the integral formulae of \PTN, \PTbd\ are universal.

\medskip
The following  relations  follow from the pentagon identity, 
\eqn\pconL{\eqalign{
 G_{ci}\left[ {j\atop b}{k \atop a}\right] &=  G_{b k^*}\left[ {i^*\atop a}{j \atop c}\right] {G_{c Q}\left[ {k^*\atop a}{k \atop a}\right]\over G_{b Q}\left[ {i^*\atop a}{i \atop a}\right] 
 }
 =G_{b^*k^*}\left[ {j\atop c^*}{i^* \atop a^*}\right] {d_i\over d_k}\cr
 &=G_{ic^* }\left[ {k^*\atop j}{a^* \atop b}\right] {d_i\over d_c} 
 }}
 where in the second equality we have used  the (Coulomb gas)  values \sgaq, particular for the chosen gauge, and the third equality is obtained repeating the first
 one. 
This relation is derived alternatively by  using \relbc\ and the cyclic symmetry of the   boundary 3-point coefficients in the l.h.s. of \reflC. 
 In particular \pconL\ implies
\eqn\lzv{
 G_{0i}\left[ {j\atop j}{k \atop k^*}\right] =
 {G_{0 Q}\left[ {k^*\atop k^*}{k \atop k^*}\right]\over G_{j Q}\left[ {i^*\atop k^*}{i \atop k^*}\right]}=
 G_{Qi}\left[{k \atop k} {j\atop j^*}\right] =
  {d_i\over d_k}\,.
}
Here we  have  replaced delta function singularities on both sides with the  residue values;
for the precise details of  treatment of these  singularities  see Appendix B of \PTbd. 

In all these relations it is assumed that the triples of representations are consistent with the corresponding fusion multiplicities.
With this data the analogs of the first two steps \derrat, \derratb\ in the rational case
now   read
  \eqn\fstL{\eqalign{
   &G_{\b\g}\left[ {\a_4\atop \a_3^*}{\a_1 \atop \a_2}\right]  G_{\b\d}^*\left[ {\a_4\atop \a_3^*}{\a_1 \atop \a_2}\right] =
   G_{\a_1\a_3^*}\left[ {\a_2\atop \b}{\g \atop \a_4^*}  \right]   G_{\a_4^*\a_2}\left[ {\a_3^*\atop \b}{\d^* \atop \a_1} \right] {d_\g d_\d \over d_{\a_3} d_{\a_2}}\cr
  &= {d_\g d_\d \over 
   d_y} \int  d y\, G_{ \a_3^* y^*}\left[ {\g^*\atop \a_2}{\d \atop \a_2}\right]   G_{\a_4^*  y}\left[ {\g\atop \a_1}{\d^* \atop \a_1}\right] \,   G_{y^* \b}\left[ {\a_2\atop \a_2}{\a_1 \atop \a_1^*}\right] G_{\a_1^* Q}\left[ {\b\atop \a_2}{\b^* \atop \a_2}\right] 
  }}
 \medskip 

\subsec{\bf The  torus identity and its application}


The basic MS torus identity in the Liouville theory is an integral relation
\eqn\tid{\eqalign{
&S_{r x}(s)\,  \int dm\, 
e^{2  \pi i (\triangle({x})-\triangle(m))}G_{s m}\left[ {x \atop x} {j_2\atop j_1}\right] G_{m p}\left[ {j_1\atop x}{j_2 \atop x}\right] \cr
&=e^{i \pi (\triangle(p) -\triangle({j_1})-\triangle({j_2}))}
\int dq\, S_{q x}(p)G_{s q}\left[ {r\atop r}{j_2 \atop j_1}\right] G_{r p}\left[ {j_2\atop q}{j_1 \atop q}\right] 
}}
The identity  is gauge invariant and can be rewritten in terms of the F matrix with 
\eqn\sgauge{
 {}^{(F)} S_{\a  x}(s)=  S_{\a x}(s){N(\a,s,\a)\over N(x, s,x)}=  {}^{(F)} S_{\a  x}(Q-s)\,.
}
It depends  on the range of the representations, here symbolically written in general form; in   the FZZ case  the notation  $\hat{S}_{\a \b} (p)$ will be used. 
  Setting  $r=0=s$,  hence $q=j_2=j_1^*=\a$,  one  obtains  expressions for the 1-point modular matrices,
  integral in the generic case.  
In the case of our main interest, when $x$ (and hence $p$)  in \tid\ is degenerate $x=x_{m,n}$,  the first integral  in \tid\  is replaced by  a finite sum with the Coulomb gas expressions 
of  the $G$ matrices  appearing, 
 \eqn\toraa{\eqalign{
S_{\a x}(p)&=
{S_{0 x}\over G_{0p}\left[ {\a\atop \a}{\a^* \atop \a}\right] } 
\sum_{u}\, e^{i \pi (2(\triangle(x)\!+\!\triangle(\a)\!-\!\triangle(u))-\triangle(p))}G_{Q u}\left[ {\a\atop \a}{x \atop x^*}\right] G_{u p^*}\left[ {x^*\atop \a}{x \atop \a}\right] \cr
&=
{S_{j0}\over G_{pQ}\left[ {x^*\atop x^*}{x \atop x^*}\right] } 
\sum_u\, e^{i \pi (2(\triangle(x)\!+\!\triangle(\a)\!-\!\triangle(u))-\triangle(p))}G_{p u }\left[ {\a\atop \a}{x \atop x^*}\right] G_{u 0}\left[ {x^*\atop \a}{x \atop \a}\right] \cr
}}
(where $u=\a+kb+l/b$ with the range of the integers $k,l$ restricted  by the fusion rule  of  the degenerate $x_{m,n}$).
The second equality follows from the pentagon identities. Note that  
$
S_{\a x}(Q)= {d_{x}\over d_{\a}} S_{\a x}\,
$ from \sgauge. 
As in the rational case,  \toraa\  
provides for $p=0$ an alternative formula for the  modular matrix \modmZ, which can be also  written in integral form
\eqn\scha{
S_{\a x_{m,n}}= S_{00}\int d\g  e^{2 \pi i (\triangle(\a)+\triangle(x_{m,n})-\triangle(\g))} \, d_\g\,  \N_{\a x_{m,n}}{}^\g
}
This and the analogous formula  for $\hat{S}_{\a\b}/S_{00}$  with $ \N_{\a x_{m,n}}{}^\g
$ replaced by  $\hat{\N}_{\a \b}{}^\g$ can be checked 
 as in the rational case.
In  the generic  integral  analog of \toraa\ 
 $\hat{S}_{\a \b} (p)$  stands in the l.h.s. of \toraa\  and an alternative representation can be  obtained similarly to
   the computation in \hosomichi\  of  the FZZ bulk-boundary constant, by solving    the   set of finite difference equations  obtained when  setting  $j_2=-b/2$ in the integral analog of \toraa, equivalent to the basic identity   \tid; see also below.

 The expression \toraa\ simplifies   if $2\a=p^*$, or $p$ (or if $2x=p$, or $p^*$)
 since,  as discussed above, the $G$ matrix in the r.h.s is simple, does not involve a summation and  reduces to a product of gauge factors. 
The simplest examples are provided taking  
\toraa\   for $x=x_{2,1}=-b/2$ and $ p=-b$, or $p=Q+b$. For these values   the sum contains two terms, $u=\a\pm b/2$,  $e^{-i \pi \triangle(-b)}=- e^{ i \pi 2 Q b}$ and 
one computes the 1-point modular matrices using the fundamental $G$ matrices \ffg,  
\sgaq, \lzv
\eqn\tmod{\eqalign{
&{{S}_{\a, - \frac{b}{2}}(-b)\over S_{0 0} }=
{d_{\a}\over d_{-b}  }
 {2i\,   e^{i\pi Q b}\,  \sin \pi 2 b\a \, \sin \pi b(2 \a-2Q) \over  \sin \pi b^2}\,, \cr
&{{S}_{\a, - \frac{b}{2}}(Q+b)\over S_{0 0} }=
{\fqd}\  {e^{i \pi Qb} \, 2i\,  \sin \pi b^2}\,.
}}

Next we set $p=0$ in \tid\ and apply  \toraa\ for the sum in the l.h.s. of \tid, using also \sgaq, or, changing notation, $ r=\a_1\,, \a=\a_2\,, q=\b$, we get
 \eqn\btot{\eqalign{
 & \int d \b  \, 
 {S_{x \b}\over S_{00}} G_{y^* \b}\left[ {\a_2\atop \a_2}{\a_1 \atop \a_1^*}\right]    G_{\a_1^* Q}\left[ {\b\atop \a_2}{\b^* \atop \a_2}\right] 
 ={S_{\a_2 x}(y^*)\over S_{0 0}}{e^{i\pi \triangle(y)}\over d_x^2}{S_{\a_1 x}(y)\over S_{0 0}}\cr
}}
This relation is what we need  when evaluating   the analog of \derrat. Indeed  we have an additional factor
$S_{\b 0}$,  coming from the measure in \loc,  which cancels the denominator of the defect eigenvalue
${S_{x\b}\over S_{0 \b}}$. Combining with \fstL\ we finally obtain
\eqn\derL{\eqalign{
& A_{\g, \d}^{(x)}=\int d\b  {S_{x\b}\over S_{0 0}} G_{\b\g}\left[ {\a_4\atop \a_3^*}{\a_1 \atop \a_2}\right]  G_{\b\d}^*\left[ {\a_4\atop \a_3^*}{\a_1 \atop \a_2}\right]  \cr
 &={d_{\g } d_\d} \sum_y G_{ \a_3^* y^*}\left[ {\g^*\atop \a_2}{\d \atop \a_2}\right] {S_{\a_2 x}(y^*)\over S_{0 x}}{e^{i\pi \triangle(y)}\over d_y}{S_{\a_1 x}(y)\over S_{0 x}}  G_{\a_4^*  y}\left[ {\g\atop \a_1}{\d^* \atop \a_1}\right] =: d_\g \sum_y  B_{\g, \d}^{(x)}(y)
}}
We have replaced the integral in \fstL\  by a sum once again using the fact  that for degenerate representations 
the  fusion matrices  are represented by   residues of the initial singular expressions. 
Using that
$S_{\a x}(Q)= {d_{x}\over d_{\a}} S_{\a x}$ and \sgaq\ one checks that for trivial defect  $x=0$ (hence $y=0$) \derL\ reduces to 
$ A_{\g, \d}^{(0)}=d_\g  \delta(\g-\d)$  in agreement with  \loc. 
\medskip

For $x$ in the continuous series, 
  the integral analog of \derL\ holds, 
with the ratio  ${\hat{S}_{x \b}\over S_{00}}$ in the l.h.s. (confirmed  \Sar\ as a defect eigenvalue), 
while  $\hat{S}_{\a_2 x}(y^*)/S_{0x} \,, \hat{S}_{\a_1 x}(y)/S_{0x}$  will appear in the r.h.s.; the remaining  q-dimension factors are unchanged.

The generic analog of   \btot\  
 considered as an expression for $\hat{S}_{\a_2 x}(y^*)$  simplifies  for the  choice $ \a_1=y/2$
 of the other charge
by the mechanism discussed above: this is more transparent in the transposed 
 version of \btot\  obtained using the identities \pconL\ and 
$
\hat{S}_{\a  x}(y)= {d_x\over d_{\a}}\hat{S}_{x \a}(y^*)\,.
$
On the r.h.s. appears the constant  $ Res_{2\a_1=y}\hat{S}_{x  \a_1 }(y^*)$, while with this choice the $G$ matrix 
in the l.h.s. of \btot\  is  replaced by a Coulomb gas correlator  from \relbc: 
\eqn\simplfB{\eqalign{
G_{y^*,  \b}\left[ {\a_2\atop \a_2}{\frac{y}{2} \atop Q- \frac{y}{2}}\right]  &  G_{Q-\frac{y}{2}, Q}\left[ {\b\atop \a_2}{\b^* \atop \a_2}\right] ={1\over d_y} G_{ \b,  y}\left[ {Q-\frac{y}{2} \atop \a_2}{\frac{y}{2} \atop \a_2}\right]  
\cr
&= { S_{ 0  0}\over   S_b^2(y)}{ S_b(\b +\frac{y}{2}-\a_2) S_b(\b +\frac{y}{2}+\a_2-Q)\over
S_b(\b -\frac{y}{2}+\a_2) S_b(\b -\frac{y}{2}-\a_2+Q) }\,.
}}
The transposed version of \btot\  (taken with either of the two modular matrices $\hat{S}_{x\b}$ or $S_{x\b}$)
is to be compared, 
with the corresponding Cardy-Lewellen type equation for the bulk-boundary reflection coefficient.
Special cases of   this equation 
have been used in \FZZ, \ZZ\ for the determination of the  half-plane bulk  1-point  
functions in the two Liouville cases. The general  equation has been exploited  in \BPb\   to give an alternative derivation of   the 
  FZZ bulk-boundary constant  $ R_x(\a, y)$, computed in   \hosomichi.
  In obtaining  \simplfB\  here   we have  followed a similar   argument to that given in \BPb\  - indeed  one recognises   in the expression 
 \simplfB\    the Fourier transform $ \tilde{R}(\a_2, y; \b -Q/2)$  
 up to $\beta$-independent factors. 
The resulting expression  for $ \hat{S}_{x\a}(y)$  can be written in terms of
 $b$-deformed hypergeometric functions (see  \PTNa, \Uen)
 \eqn\formF{\eqalign{
& \hat{S}_{x\a}(y)e^{i\pi  \triangle(y)/2}=  {d_\a\over d_x}\hat{S}_{\a x}(y^*)e^{i\pi  \triangle(y)/2}\cr
&=\sum_{\pm} {  S_b(\pm(2\a\!-\!Q)\!+\!y)\over S_b(\pm(2\a\!-\!Q))}e^{i \pi (\pm(2\a\!-\!Q)\!+\!y)(2x\!-\!Q)}
F_b(y, \pm(2\a\!-\!Q)\!+\!y; \pm(2\a\!-\!Q)\!+\!Q; 2x\!-\!Q)\cr
&=
 {S_b(2\a+y-Q)\over S_b(2\a-Q)}(\sum_{\pm} e^{\pm i \pi (2\a+y-Q)(2x-Q)}
F_b(y, 2\a+y-Q; 2\a; \pm(2x-Q))\,.
}}
For $y=0$ it reproduces $\hat{S}_{x\a}$ in \modF\ and the residue  of $\hat{S}_{x \a}(y^*)$  at $\a=y/2$ 
is consistent with  the FZZ analog  of eqn  \btot\  
at this value. The required symmetries like the one in the first line of \formF\ are checked exploiting 
relations for the $b$-deformed   hypergeometric functions.
Note that 
the  ratio   
$${\hat{S}_{ x \a}(y)e^{i\pi  \triangle(y)/2}\over W(\a) g(\a, y,\a)}= {{}^F\hat{S}_{ x \a}(y)e^{i\pi  \triangle(y)/2}\over W(\a) g(x, y,x)}$$  
\noindent 
has the correct properties under reflections with respect to the bulk $\a$ and the boundary $y$ field  charges as  required for the  FZZ bulk-boundary constant  $ R_x(\a, y)$.

For 
(odd) degenerate $y=y_{2m+1,2n+1}=
-mb-n/b$,  one furthermore obtains, 
accounting for the residues of the poles of the integrand defining the hypergeometric functions
a formula for degenerate $x=x_{r,s}$: the computation is analogous to the
 derivation of the ZZ bulk-boundary coefficient in \BPb. 
Namely, denoting $\g_{k,l}= kb +l/b$, 
  \eqn\formFdeg{\eqalign{
& {S}_{x_{r,s}\a}(y)={S}_{\a x_{r,s}}(y^*){d_\a\over d_{x_{r,s}}}= \hat{S}_{x_{r,s} \a}(y)-\hat{S}_{x_{-r,s}\a}(y)
=
e^{-{i\pi  \triangle(y)\over 2}}\, 
\sum_{k=0}^m\sum_{l=0}^n{S}_{x_{r,s}\,, \g_{k,l}\!+\!\frac{y}{2}\!+\!\a} \, \times \cr
& {(-1)^{kn +ml}S_b((m+1)b) S_b(\frac{n+1}{b})
\over
S_b((m+1-k)b) S_b(\frac{n+1-l}{b}) 
S_b((k+1)b)  S_b(\frac{l+1}{b}) } {S_b(2\a) \over  S_b(2\a\!+\!\g_{k,l}) } {S_b(2Q -2 \a)  \over S_b(2Q\!-\!y\!-\!\g_{k,l}\!-\!2\a) }\,.
}}
The values of $y$  are  restricted by the degenerate 
 fusion multiplicity,  
    $\!0\!\le\!m\le\!r\!-\!1$, $0\!\le\!n\!\le s\!-\!1$. 
The formula  \formFdeg\   provides  alternative  
to  \toraa\ representation for $ {S}_{\a x_{r,s}}(Q-y_{2m+1,2n+1})$. It  is checked  to reproduce the particular example in \tmod.


\subsec{\bf The 't Hooft operator: example}


Here we compute \derL\ for    the simplest example considered in \AGGTV, \DGOT, namely $x=-b/2$, 
so that    $y$ takes the values $y=0, -b$.

The $G$ matrices for $\d=\g\pm b$  are all straightforward to compute, 
  being related as explained above,  to trivial boundary OPE coefficients, 
\eqn\Gp{\eqalign{
G_{\s_4, Q\!+\!b}\left[\matrix{Q\!-\!\g\!&\!\g\!+\!b\cr \s_3\!&\!\s_3} \right]&\!=\!-{d_{-b/2} d_{\s_3} \over d_{\g+b}}{\sin \pi 2  \s_3 b\,  \sin \pi b (2\s_3-2Q) \over  \sin \pi 2 \g b\,  \sin \pi b(2\g-Q)}G_{\s_4,\!-b}\left[\matrix{Q\!-\!\g\!&\!\g\!+\!b\cr \s_3\!&\!\s_3} \right] \cr
&\!=\!{d_{-b/2} d_{-b}\over d_{\g+b}}{\sin \pi b(\g-\s_3+\s_4)\sin \pi b(\g-\s_4+\s_3)\over  \sin \pi  2\g b\,  \sin \pi b(2\g-Q)}\,,\cr
&{}\cr
G_{\s_4,Q\!+\!b}\left[\matrix{Q\!-\!\g\!&\!\g\!-\!b\cr \s_3\!&\!\s_3} \right]&\!=\!-{d_{-b/2} d_{\s_3} \over d_{\g+b}}{\sin \pi 2  \s_3 b\,  \sin \pi b (2\s_3-2Q) \over  \sin \pi b(2\g\!-\!2Q)\,  \sin \pi b(2\g\!-\!Q)}G_{\s_4,\!-b}\left[\matrix{Q\!-\!\g\!&\!\g\!-\!b\cr \s_3\!&\!\s_3} \right] \cr
&\!=\!{\fqd d_{-b }\over d_{\g-b}} {\sin \pi b(\s_3+\s_4-\g)\,\sin \pi b(2Q-\g-\s_4-\s_3)\over
 \sin \pi b(2\g\!-\!2Q)\, \sin \pi b(2\g\!-\!Q)}\,.
}}   
  For the  only matrix that involves two terms   
(the case $m=1\,, n=0$ in \PTR) one gets  
\eqn\Gno{\eqalign{
G_{\s_4,\!-\!b}\left[\matrix{Q\!-\!\g\!&\!\g\cr \s_3\!&\!\s_3} \right]&\!=\!-{d_{-b}\over d_{\s_3 }}
{\cos \pi b(2\g\!-\!Q)  \cos \pi b(2\s_3\!-\!Q) +\cos \pi b(2\s_4\!-\!Q)  \cos \pi b^2
\over   \sin \pi  2\s_3 b\,   \sin \pi b(2\s_3 -2Q)}\cr
&=G_{\s_4,Q\!+\!b}\left[\matrix{Q\!-\!\s_3\!&\!\s_3\cr \g\!&\!\g} \right] \,.
}}
Besides \tmod\ we also need 
\eqn\Gs{
{S_{\a_1, -\frac{b}{2}}\over S_{\a_1, 0}}= 2\cos \pi b(2\a_1-Q)\,, \  {S_{\a_2, -\frac{b}{2}}(Q)\over S_{00}} ={\fqd\over d_{\a_2}} {S_{\a_2, -\frac{b}{2}}\over S_{00}} = 
\fqd 2 \cos \pi b(2\a_2-Q)\,. 
}
Let us change the notation 
$(\a_3,\a_2,\a_1, \a_4) \to (\s_4,\s_3,\s_2, \s_1)$, 
so that $B_{\g,\d}^{(x)}(y)$,   as defined in \derL, reads
\eqn\derLa{
 B_{\g, \d}^{(x)}(y)={d_\d\over d_y d_x^2} G_{ \s_4^* y^*}\left[ {\g^*\atop \s_3}{\d \atop \s_3}\right] {S_{\s_3 x}(y^*)\over S_{0 0}}{e^{i\pi \triangle(y)}}{S_{\s_2 x}(y)\over S_{0 0}}  G_{\s_1^*  y}\left[ {\g\atop \s_2}{\d^* \atop \s_2}\right]\,.
 }
Collecting all formulae obtained from \Gp, \Gno\   with the proper conjugations and change of variables,
and using also \sgaq,  \tmod,  \Gs,  we obtain for  \derLa:
\eqn\prpgbma{\eqalign{
&B_{\g,\g-b}^{(-b/2)}(-b)\!=\!-
{4\sin \pi b(Q\!+\!\s_3^4\!-\!\g)\sin \pi b(Q\!+\!\s_4^3\!-\!\g)
 \sin \pi b(2Q\!-\!\g\!-\!\s_{12}\!)\sin \pi b(\s_{12}\!-\!\g\!)\over 
 \sin \pi b (2\g-Q)\,  \sin \pi b (2\g-2Q)} 
\cr
&B_{\g,\g+b}^{(-b/2)}(-b)\!=\!-
\, {4\sin \pi b(Q\!+\!\g\!-\!\s_{34})\sin \pi b(\g\!+\!\s_{34}\!-\!Q)
\sin \pi b(\g\!+\!\s_2^1)\sin \pi b(\g\!+\!\s_1^2)
\over 
 \sin \pi b (2\g-Q)\,  \sin \pi b  2\g} 
 }}
 \eqn\prpgbm{\eqalign{
&B_{\g, \g}^{(-b/2)}(-b)
=
\, {\cos \pi b(2\g-Q)  \cos \pi b(2\s_3\!-\!Q) -\cos \pi b(2\s_4\!-\!Q)  \cos \pi b^2
\over  \fqd\,  \sin \pi b 2\g  \sin \pi b(2\g-2Q)} \, \times \cr
&
\qquad \qquad \qquad \ \ 4 (\cos \pi b(2\g\!-\!Q)  \cos \pi b(2\s_2\!-\!Q) -\cos \pi b(2\s_1\!-\!Q)  \cos \pi b^2)
\cr
&{}\cr
&B_{\g, \g}^{(-b/2)}(0)=
{4 \cos \pi b(2\s_3-Q)\,  \cos \pi b(2\s_2-Q) \over \fqd}\,, \ \  \fqd= - 2\cos \pi b^2\,.
}}
where $\s_{12}=\s_1\!+\!\s_2\,, \s_1^2=\s_1\!-\!\s_2$, etc.
The above, normalised  by $d_{-b/2}$,   expressions for $B_{\g, \d}(-p)/d_{-b/2}$  
should be compared with formulae (5.32-34) of \DGOT.
Apart from, presumably,  a sign  typo  in (5.34):
  
$\sin \pi b(\a\!-\!m_{12}\!-\!b) \sin \pi b(\a\!-\!m_{34}\!-\!b) \to
\sin \pi b(\a\!-\!m_{12}\!+\!b) \sin \pi b(\a\!-\!m_{34}\!+\!b)$

\noindent
the formulae coincide for 
$(\g; \s_4,\s_3,\s_2,\s_1) \to (\a; m_4,m_3,m_2, Q-m_1)$,
i.e., up to  a 
reflection 
of one of  the charges.  
Here $\s_1 =\a_4$ is the  charge of the first  vertex operator
  in the  Wilson loop   channel, 
cf. \gbl.

The sum of the  
two terms in \prpgbm\  can be cast   into 
a   form which makes explicit the symmetry $(\s_2,\s_3) \to (\s_4,\s_1)$.
It 
 coincides (up to a relative sign) with  the   term  $H_0$ in (5.39) of \AGGTV\  upon   identification  of the charges: $2\s_j-Q= 2i m_j\,, \, 2\g-Q=2iP$. 
 In  the initial basis of conformal blocks the duality relation reads
\eqn\drfin{\eqalign{
&\int d\b\, {C(\a_4,\a_3,\b)C(\b^*, \a_2,\a_1)  }
 {S_{x \b}\over S_{0\b}} \,  |{}^{(F)}\CG_\b(\a_4,\a_3,\a_2,\a_1; \tilde{z})|^2\cr
&=\int d\g \int d\d \sum_y  C(\a_3,\a_2, (\g,\d) ) C((\g^*,\d^*), \a_1,\a_4)\, {W(\g)\over W(\d)} B_{\g, \d}^{(x)}(y) \times\cr
&\qquad {}^{(F)}\CG_{\g}(\a_3,\a_2,\a_1, \a_4; z) {}^{(F)}\CG_{\d}^*(\a_3,\a_2,\a_1, \a_4;{ z})\,
}} 
where we have denoted (consistent with \pn\ for $\g=\d$)
 \eqn\fedstr{\eqalign{
&C(\a_3,\a_2, (\g,\d) ) C((\g^*,\d^*), \a_1,\a_4){W(\d)\over W(\g)}:=
\cr
& {(2\pi)^2  \l^{-Q\over b} \prod_{i=1}^4   
{W(Q-\a_i) } \, S_{\g0}
\over N( \a_3^*,\g, \a_2)N( \g, \a_4,\a_1)N^*( \a_3^*,\d, \a_2)N^*( \d, \a_4,\a_1)} \cr
&=C(\a_3,\a_2, \g ) C(Q-\g, \a_1,\a_4)\, 
{N(\a_3,Q-\g, Q-\a_2)N(\a_1, Q-\a_4,\g)\over N(\a_3,Q-\d, Q-\a_2)N(\a_1, Q-\a_4,\d)} 
\,.
}}
 Multiplying $B_{\g, \g\pm b}^{(-b/2)}(-b) $ in \prpgbma\ with  the    ratio of $N$-factors, relative to the diagonal constant    in the last line in \fedstr, we get expressions  invariant under any reflection 
 $\a_i\!\to\!Q-\a_i$ of the four charges. These normalised expressions   coincide  
with $H_{\pm}$ in (5.38) of \AGGTV\  (up to an 
overall factor $2\pi$)  under the above identification of the charges.
 

 \appendix{A}{The  defects and the  OPE coefficients of local fields}
 
 
  In the non-diagonal rational cases  the identity contribution in the duality relation for  the correlators 
  \vone,\voneb\  is nontrivial and 
implies  an explicit formula \PZ\  for the relative OPE coefficients of  local  fields $ \Phi_{I;\za}(z,\bz)\,, I=(i,\bar{i})\,, \a=1,..., Z_{i \bar{i}}$  of  arbitrary integer spin
  \eqn\locop{
 \Phi_{I;\za}(z,\bar z)= \sum_{j,\bj,k,\bk,\zb,\zg,t,\bar{t}}
\ d_{(I; \za) (J; \zb)}^{(K; \zg); t,\bar{t} }\
\Big(\phi_{ij;t}^k(z)\otimes
\phi_{\bi \bj; \bar{t}}^{\bk}(\bar z)\Big)_{\za \, \zb}^{\zg}\,. 
}
 Namely  (restricting to the  $sl(2)$ case)  one  obtains
  \eqn\oped{\eqalign{
\sum_{k,\bar{k}, \g, \g'}\,
d_{(I^*; \a) (J^*; \b)}^{(K^*; \g) }\, d_{(I; \a') (J; \b')}^{(K; \g') }
{\Psi_x^{(K;\g,\g')}\over \Psi_x^{(1)}}&={\Psi_{x}^{(I;\a,\a')}\over \Psi_x^{(1)}}
\, {\Psi_x^{(J;\b, \b')}\over \Psi_x^{(1)}} \cr
}}
Using the unitarity of $\Psi$ one gets an expression  for the product of OPE coefficients
which involves a summation over the complete  set of defects
for the given modular invariant.\foot{We use the opportunity to correct some inaccuracies in \PZ: 
 formula \oped\ is slightly  more general than what stated in \PZ\  and  reproduces also some of the signs in the case with non-commutative $\tN$, the $D_{\rm even}$ series:  
 note that there are different bases  for the pair of doubled fields in this case 
 and accordingly different bases for the $\Psi$ matrices. } The $\Psi$-ratios in \oped\  serve as 1-dimensional representations
of an associative, commutative algebra, dual to the fusion algebra of defects. This universal algebra    
generalises  the Pasquier algebra \PasM\
associated with each of the ADE    \nimreps\ in \nmrep,  which  determines 
the subset of   OPE coefficients with  scalar labels  only \PZa.

\bigskip \bigskip
\noindent
{\bf Acknowledgements}
\medskip
\noindent

\noindent
It is a pleasure to thank   Jean-Bernard  Zuber  for the  interest in this work and   numerous  
useful discussions.  I would also like to   thank   Karl-Henning Rehren for the invitation    
 to give a series of lectures  on 2d CFT at  the  Institut   f\"ur Theoretische Physik,   
 Universit\"at  G\"ottingen  - the   initial motivation for this    work  grew out of that effort. 
 The   hospitality of  the  Service de Physique Th\'eorique, CEA-Saclay and   LPTHE, 
Universit\'e  Pierre et Marie Curie,  Paris,  is  also acknowledged. This research is  
supported   by the  Bulgarian NSF grant {\it DO 02-257}.

\listrefs

\bye